\documentclass[aps,prl,
reprint,
amsmath,amssymb,
]{revtex4-1}

\usepackage{subfigure}
\usepackage{graphicx}
\usepackage{epstopdf}
\usepackage{dcolumn}
\usepackage{bm}

\preprint{APS/123-QED}

\begin{document}

\title{Self-Amplification of Coherent Energy Modulation in Seeded Free-Electron Lasers}

\author{Jiawei Yan$^{1,2}$, Zhangfeng Gao$^{1,2}$, Zheng Qi$^{3}$, Kaiqing Zhang$^{3}$, Kaishang Zhou$^{3}$,Tao Liu$^{3}$, Si Chen$^{3} $, Chao Feng$^{3}$, \\ Chunlei Li$^{3}$, Lie Feng$^{3}$, Taihe Lan$^{3}$, Wenyan Zhang$^{3}$, Xingtao Wang$^{3}$, Xuan Li$^{3}$, Zenggong Jiang$^{3}$, \\ Baoliang Wang$^{3}$, Zhen Wang$^{3}$, Duan Gu$^{3}$, Meng Zhang$^{3}$, Haixiao Deng$^{3}$} \email{denghaixiao@zjlab.org.cn} \author{Qiang Gu$^{3}$, \\   Yongbin Leng$^{3}$, Lixin Yin$^{3}$, Bo Liu$^{3}$, Dong Wang$^{3}$} \author{Zhentang Zhao$^{3}$}

\email{zhentangzhao@zjlab.org.cn}
\affiliation{%
	$^1$Shanghai Institute of Applied Physics, Chinese Academy of Sciences, Shanghai 201800, China\\ $^2$University of Chinese Academy of Sciences, Beijing 100049, China\\ $^3$Shanghai Advanced Research Institute, Chinese Academy of Sciences, Shanghai 201210, China. 
}%


\begin{abstract}

The spectroscopic techniques for time-resolved fine analysis of matter require coherent X-ray radiation with femtosecond duration and high average brightness. Seeded free-electron lasers (FELs), which use the frequency up-conversion of an external seed laser to improve temporal coherence, are ideal for providing fully coherent soft X-ray pulses.  However, it is difficult to operate seeded FELs at a high repetition rate due to the limitations of present state-of-the-art laser systems. Here, we report a novel self-modulation method for enhancing laser-induced energy modulation, thereby significantly reducing the requirement of an external laser system. Driven by this scheme, we experimentally realize high harmonic generation in a seeded FEL using an unprecedentedly small external laser-induced energy modulation. An electron beam with a laser-induced energy modulation as small as 1.8 times the slice energy spread is used for lasing at the 7th harmonic of a 266-nm seed laser in a single-stage high-gain harmonic generation (HGHG) setup and the 30th harmonic of the seed laser in a two-stage HGHG setup. The results mark a major step towards a high-repetition-rate, fully coherent X-ray FEL.

\end{abstract}

\maketitle

X-ray free-electron lasers (FELs) that provide high-brightness pulses of femtosecond duration have enabled new research in various scientific fields \cite{Pellegrinireview, feng2018review}. To date, most successful FEL-based experiments have investigated the internal structure or ordering of materials, which is compatible with single-pulse detection \cite{serafini2019marix}. In contrast, spectroscopic probes used to study magnetic and electronic structures require a higher average photon flux on the sample. Therefore, coherent FEL with a high repetition rate is required.

Most X-ray FEL facilities worldwide \cite{lcls,sacla,pal,decking2020mhz} employ the mechanism of self-amplified spontaneous emission (SASE) \cite{4}. The SASE scheme can obtain FEL pulses with sub-angstrom wavelengths but limited temporal coherence, as the initial amplification arises from the electron-beam shot noise. The phase and intensity fluctuations of the SASE scheme severely limit the use of X-ray spectroscopy. Self-seeding schemes \cite{feldhaus1997possible, geloni2011novel} can be used to achieve narrow-bandwidth pulses but at the cost of shot-to-shot intensity fluctuations. Seeded FELs \cite{yu1991generation,yu1997high,stupakov2009using} triggered by stable, coherent external lasers ensure output FEL pulses with a high degree of temporal coherence and small pulse energy fluctuation, as has been demonstrated by analytical calculations and experimental results in the ultraviolet to soft X-ray range \cite{allaria2012highly,PhysRevSTAB.16.020704,allaria2013two,zhao2012first,hemsing2016echo,ribivc2019coherent,feng2019coherent}.

In recent years, based on the superconducting linac, the XFELs with repetition rates of up to several MHz have been proposed \cite{decking2020mhz,stohr2011linac,zhu2017sclf,serafini2019marix}. A seeded FEL with such a high repetition rate can meet the requirements of high-resolution spectroscopic techniques for fine analysis of the matter. Meanwhile, borrowing the idea from seeded FELs, most storage-ring-based FELs employ external lasers to manipulate electron beams to precisely tailor the properties of the radiation pulses \cite{xiang2010generating, feng2017storage,wang2019angular}. However, owing to the limitations of present state-of-the-art laser systems, the repetition rate of an external seed laser with sufficient power to manipulate the electron beam is limited to the kilohertz range. There is a continuing trend towards higher repetition rates for future XFEL facilities and scientific requirements. 

Various methods to realize a high-repetition-rate seed source are under investigation. The concept of employing an FEL oscillator as the seeding source for subsequent cascades is proposed \cite{radiatorfirst,li2018high,petrillo2020coherent}. Recently, an optical, resonator-like seed recirculation feedback system is introduced to recirculate the radiation in the modulator to seed the following electron bunches \cite{PhysRevAccelBeams.23.071302}. These schemes require further experimental validation. In addition, for conventional laser systems, the critical limit of the repetition rate is the thermal effect of the optics, so the repetition rate can be effectively increased by reducing the peak power.

\begin{figure*}[htp] 
	\centering 
	\includegraphics[width=0.9\linewidth]{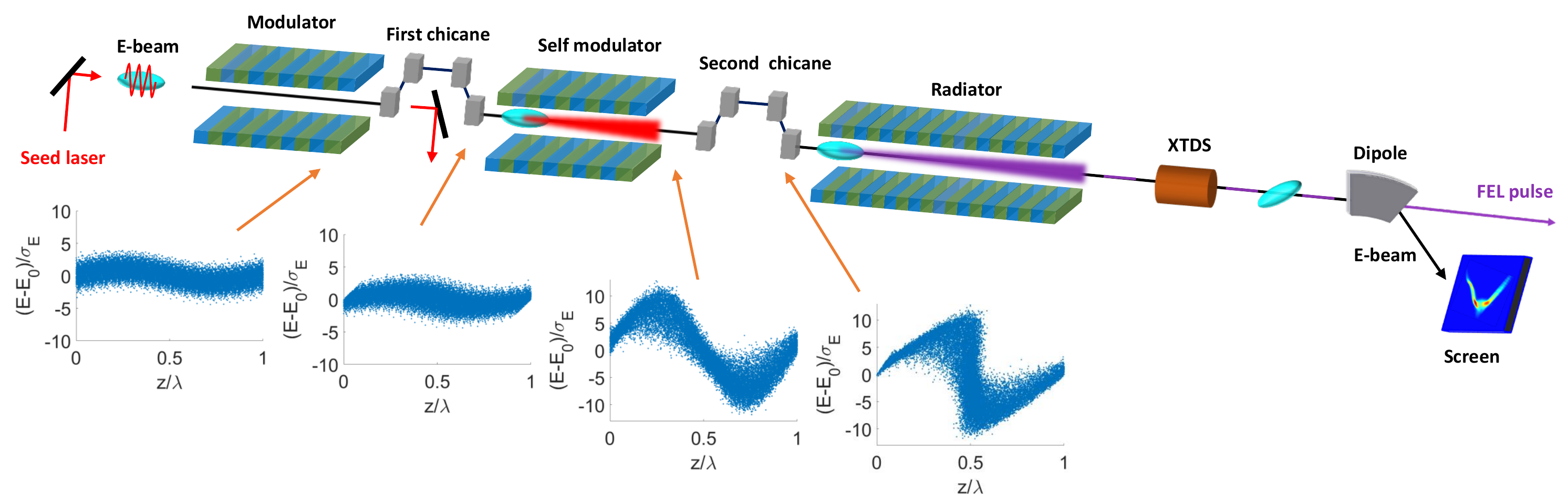}
	\caption{The self-modulation scheme together with the electron-beam longitudinal phase spaces at various positions.}
	\label{scheme01} 
\end{figure*} 


In this Letter, we report the use of self-amplification of coherent energy modulation in a seeded FEL to relax the power requirement of an external seed laser by more than an order of magnitude. The setup of this scheme is displayed in Fig.\ \ref{scheme01}. This setup is similar to the optical klystron configuration \cite{PhysRevSTAB.9.070702, PhysRevLett.114.013901} and the pre-density modulation scheme \cite{feng2010pre, feng2015theoretical} proposed to enhance the microbunching and reduce the energy spread. Compared to the single-stage high-gain harmonic generation (HGHG) \cite{yu1991generation}, an additional chicane and another short undulator are added after the modulator in this scheme. The electron beam interacts with an external laser in the modulator and is induced with an energy modulation equal to or twice the slice energy spread ($\sigma_{E}$). The first chicane in this scheme is used for density modulation of the electron beam. Because the required energy modulation amplitude for amplification at the $nth$ harmonic should be $n$-fold larger than the slice energy spread in the HGHG scheme, an electron beam with such a small energy modulation amplitude is difficult to be used for lasing at high harmonics. However, it can be used to produce strong radiation at the fundamental wavelength. Therefore, the microbunched beam is sent to the second undulator, referred to as a self-modulator, where the electron beam generates intense coherent radiation at the fundamental wavelength and is modulated by its own radiation. As indicated by the phase spaces of the electron beam in Fig.\ \ref{scheme01}, the self-modulation can significantly enhance the initial energy modulation, thus reducing the requirement for the peak power of the external seed laser.

Since the intensity of the coherent radiation generated by a microbunched beam is strongly coupled with the bunching factor and transverse size of the electron beam \cite{yu2002theory}, the self-modulation can be controlled by the intensity of the seed laser, the chicane before the self-modulator, or the transverse focusing.  Based on the amplified energy modulation, various seeded FEL schemes can be combined following the self-modulator. As presented in Fig.\ \ref{scheme01}, a second chicane and a radiator after the self-modulator, used to achieve a large bunching factor and lasing at high harmonics, make it a typical HGHG layout. 

A proof-of-principle experiment was conducted at the Shanghai Soft X-ray FEL test facility (SXFEL) \cite{zhao2017status} to demonstrate the self-amplification of coherent energy modulation. The first stage of the SXFEL has the same setup presented in Fig.\ \ref{scheme01}, which comprises a seed laser with a wavelength of 266 nm and a pulse length of 160 fs (FWHM), two modulators of length 1.5 m and period 80 mm, and two magnetic chicanes. The second modulator is treated as the self-modulator in the experiment. Connected to the second chicane, four undulator segments of length 3 m and period 40 mm are adopted as the radiator of the first stage.  

\begin{figure}[!htb]
	\centering
	\includegraphics[width=0.7\linewidth]{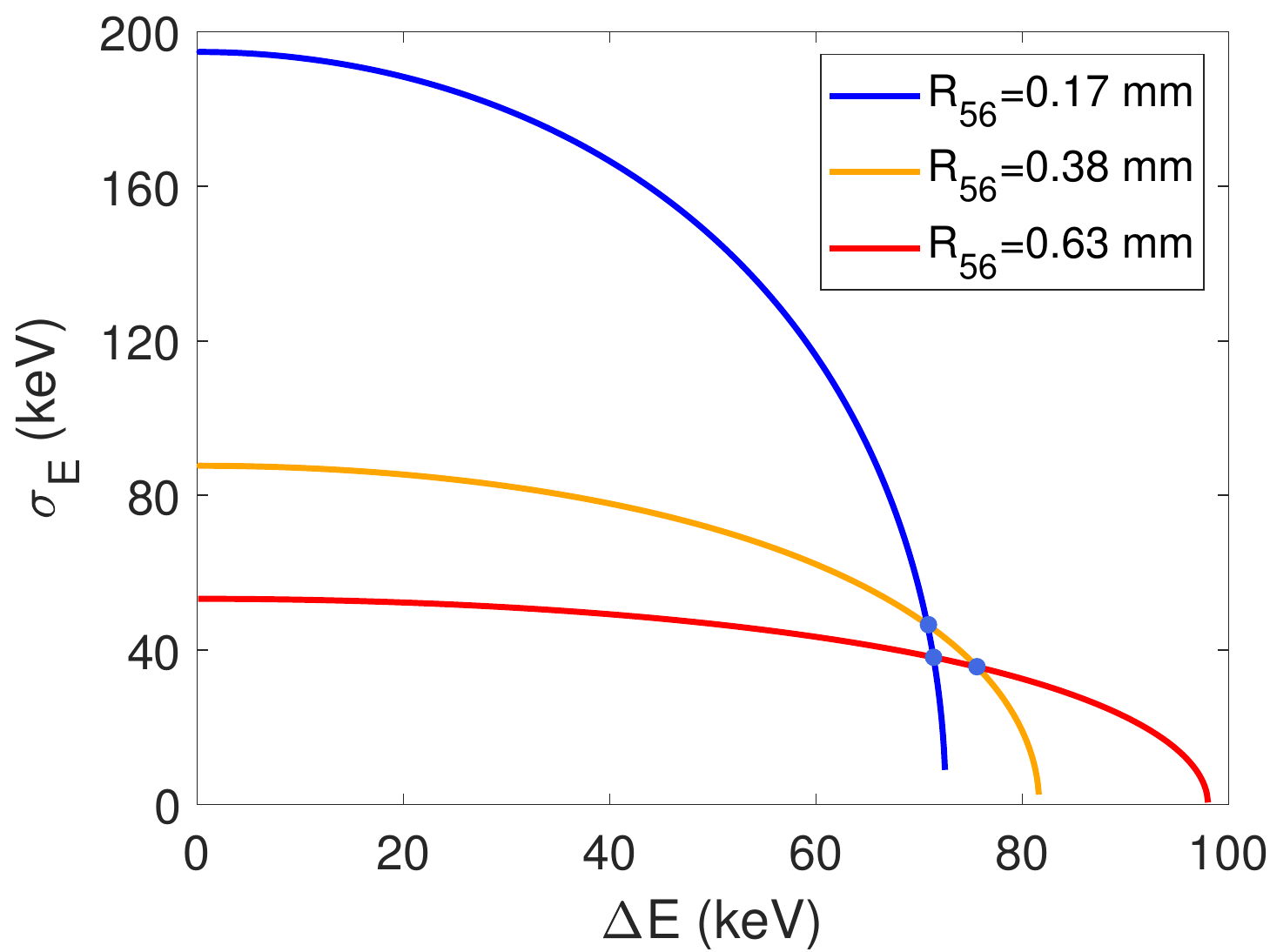}
	\caption{Calculation results of energy modulation amplitude and average slice energy spread according to the coherent radiation based method \cite{feng2011measurement}.}
	\label{r561scan}
\end{figure}

In the experiment, we first employed the coherent radiation based method \cite{feng2011measurement} to measure the laser-induced energy modulation. In the measurement, the seed laser with three different pulse energies of 38.10, 6.10, and 1.56 $\rm \mu J$ was used to interact with the electron beam with an energy of 795 MeV and a bunch charge of 550 pC in the modulator. The peak current, normalized transverse emittance, and envelope size of the electron beam are around 600 A, 1.5 $\rm mm\ mrad $, and 300 $\mu m$, respectively. The modulated electron beam was used to generate coherent radiation at the fundamental wavelength in the self-modulator. The first chicane was scanned to determine the optimal dispersion strength that maximizes the coherent radiation intensity from the self-modulator under different pulse energies of the seed laser. For each optimal dispersion strength, a numerical relationship between average slice energy spread and energy modulation amplitude can be obtained. The obtained optimal $R_{56}$ of the first chicane under the three pulse energies were 0.17, 0.38, 0.63 mm, respectively \cite{SMaterial}. Fig.\ \ref{r561scan} shows the three numerical curves obtained from the three optimal dispersion strength, where all the energy modulation amplitudes are scaled down to the amplitude induced by a 1.56 $\rm \mu J$ seed laser. The intersection of any two curves is a solution for the measurement. Thus, three solutions were obtained and the average of them was treated as the final measurement result. The measured slice energy spread and energy modulation amplitude induced by the seed laser of 1.56 $\rm \mu J$ are 40 and 73 keV, respectively. Thus, the energy modulation amplitude is approximately 1.8$\sigma_{E}$. The pulse energy of 1.56 $\rm \mu J$ was used in the following experiments to verify the feasibility of the self-modulation scheme. 

\begin{figure}[htp] 
	\centering 
	\subfigure[]{\includegraphics*[width=0.46\linewidth]{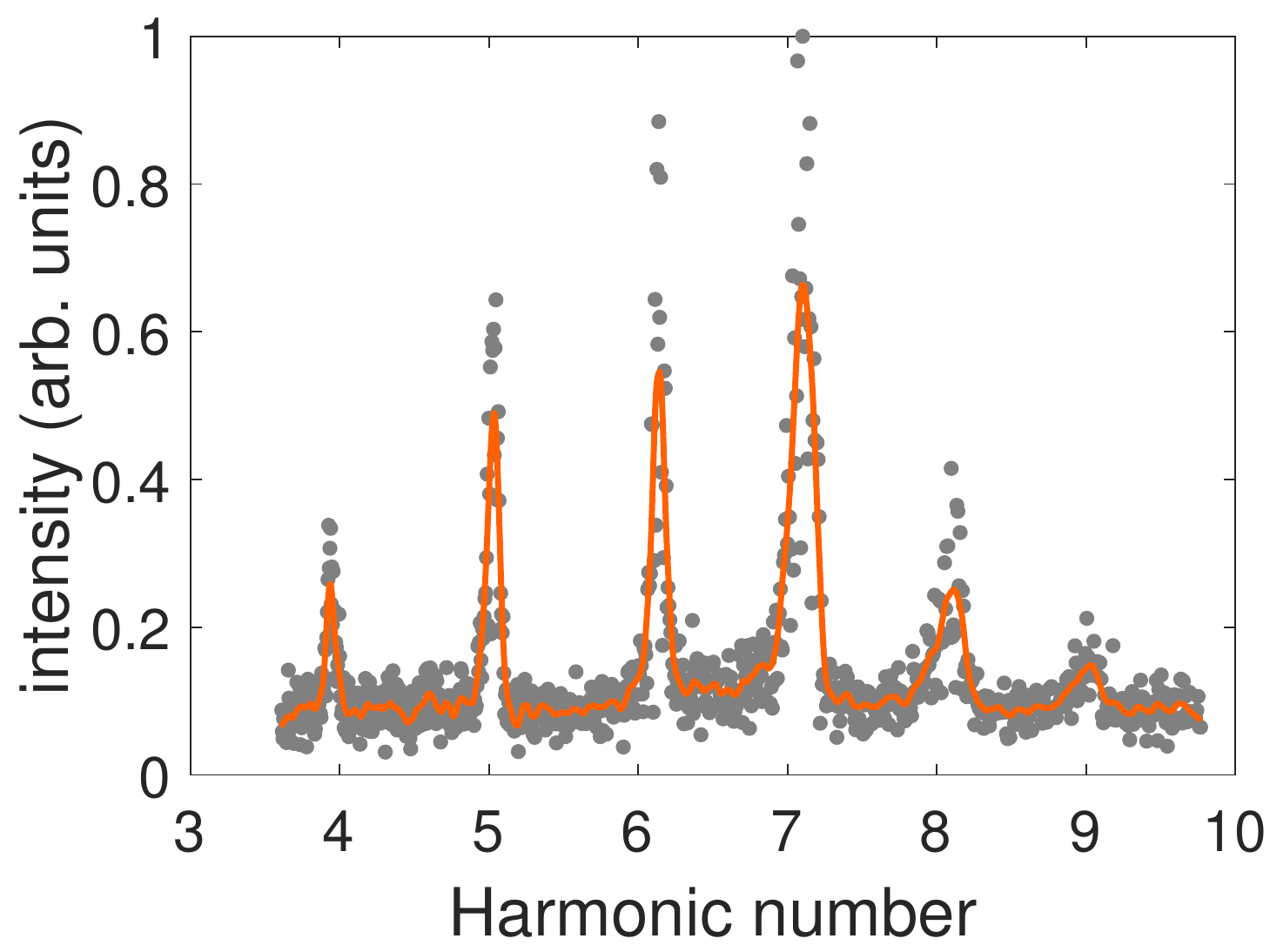}}
	\subfigure[]{\includegraphics*[width=0.45\linewidth]{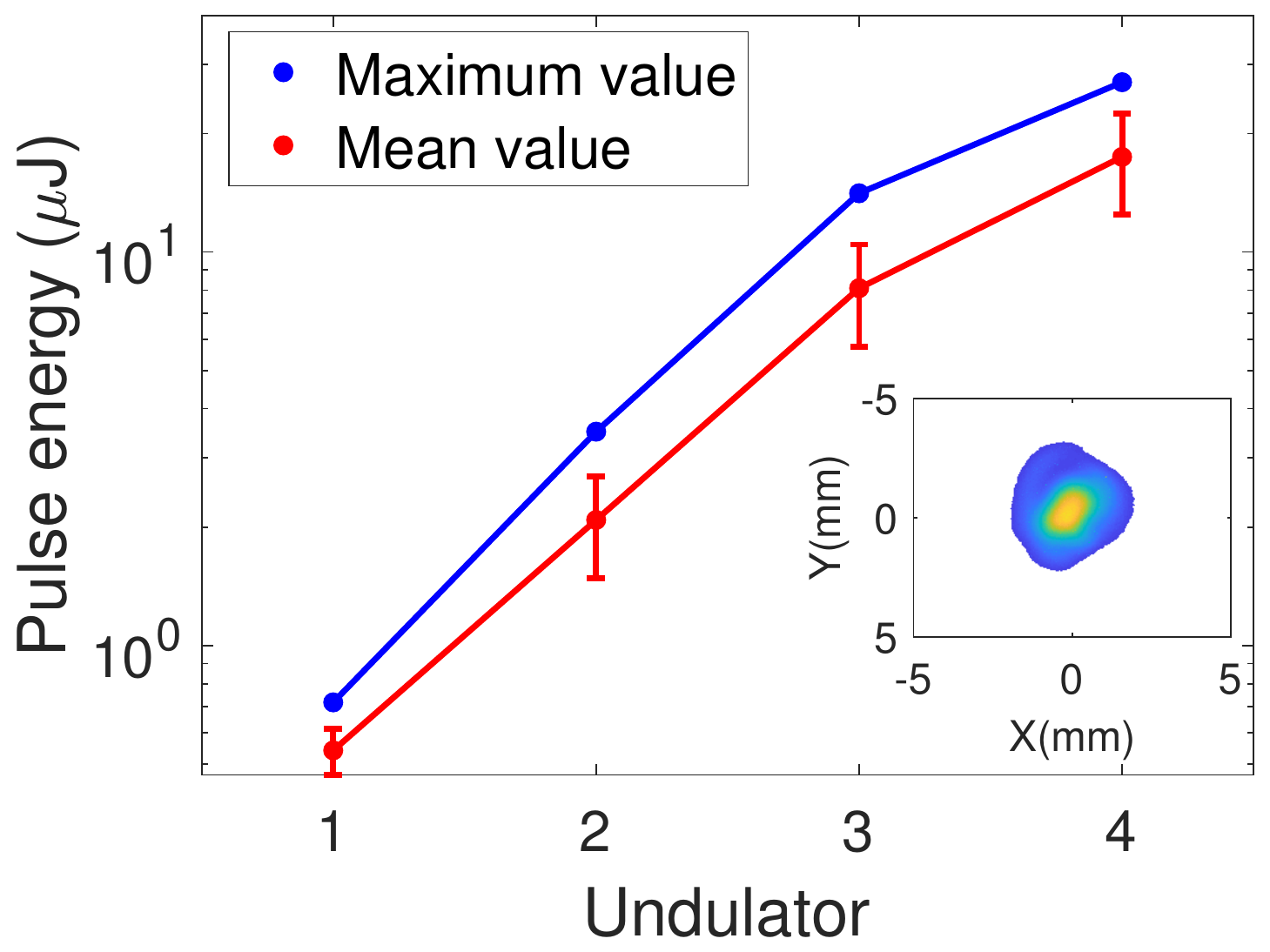}}
	\caption{(a) The measured intensities (points) of the coherent radiation at various harmonic numbers and the envelope (curve) obtained by smoothing the measured data. (b) Measured FEL gain curve at the 7th harmonic, where the dispersion strength of the second chicane is optimized. The error bars represent root mean square intensity fluctuations. Inset displays a typical measured FEL transverse profile.}
	\label{Gain} 
\end{figure}

We further performed the self-modulation and demonstrated the amplification of the laser-induced energy modulation. The $R_{56}$ of the first chicane was set at the optimal value of 0.63 mm obtained in the previous measurements. To verify the energy modulation enhancement, the electron beam was sent to the radiator for lasing at high harmonics, where only one undulator segment was used. The undulator segment gap was scanned continuously from 9.2 to 18 mm, which contained resonances at the 4th to 10th harmonics of the external laser. When the self-modulator was removed, no coherent radiation was detected by the photodiode after the radiator. Subsequently, the resonance of the self-modulator was tuned to the fundamental wavelength of the seed laser. In this case, coherent radiation can be detected even when the radiator is resonant at the 9th harmonic. The result proves the enhancement of the initial energy modulation. Fig.\ \ref{Gain} (a) displays the coherent radiation intensity at various harmonics when the $ R_{56}$ of the second chicane is set to 0.16 mm, which was not precisely optimized for a specific wavelength.

Because intense radiation can be detected at the 7th harmonic of the seed laser, i.e., 38 nm, we carefully optimized the dispersion strength of the second chicane at this wavelength. The optimal $ R_{56}$ of the second chicane for the 7th harmonic is 0.17 mm. As the energy modulation is not directly induced by the external laser, it is difficult to measure the enhanced energy modulation amplitude using the coherent radiation based method. To roughly evaluate the energy modulation amplitude, we can consider that the enhanced energy modulation is induced by a strong seed laser and the initial slice energy spread of the electron beam is the previously measured 40 keV. According to the relationship between the optimal dispersion strength and the energy modulation amplitude \cite{feng2011measurement}, the energy modulation amplitude after the self-modulation can be estimated as 218 keV. This means that the energy modulation amplitude was increased approximately threefold. 

\begin{figure}[!htb]
	\centering
	\subfigure[]{\includegraphics*[width=0.47\linewidth]{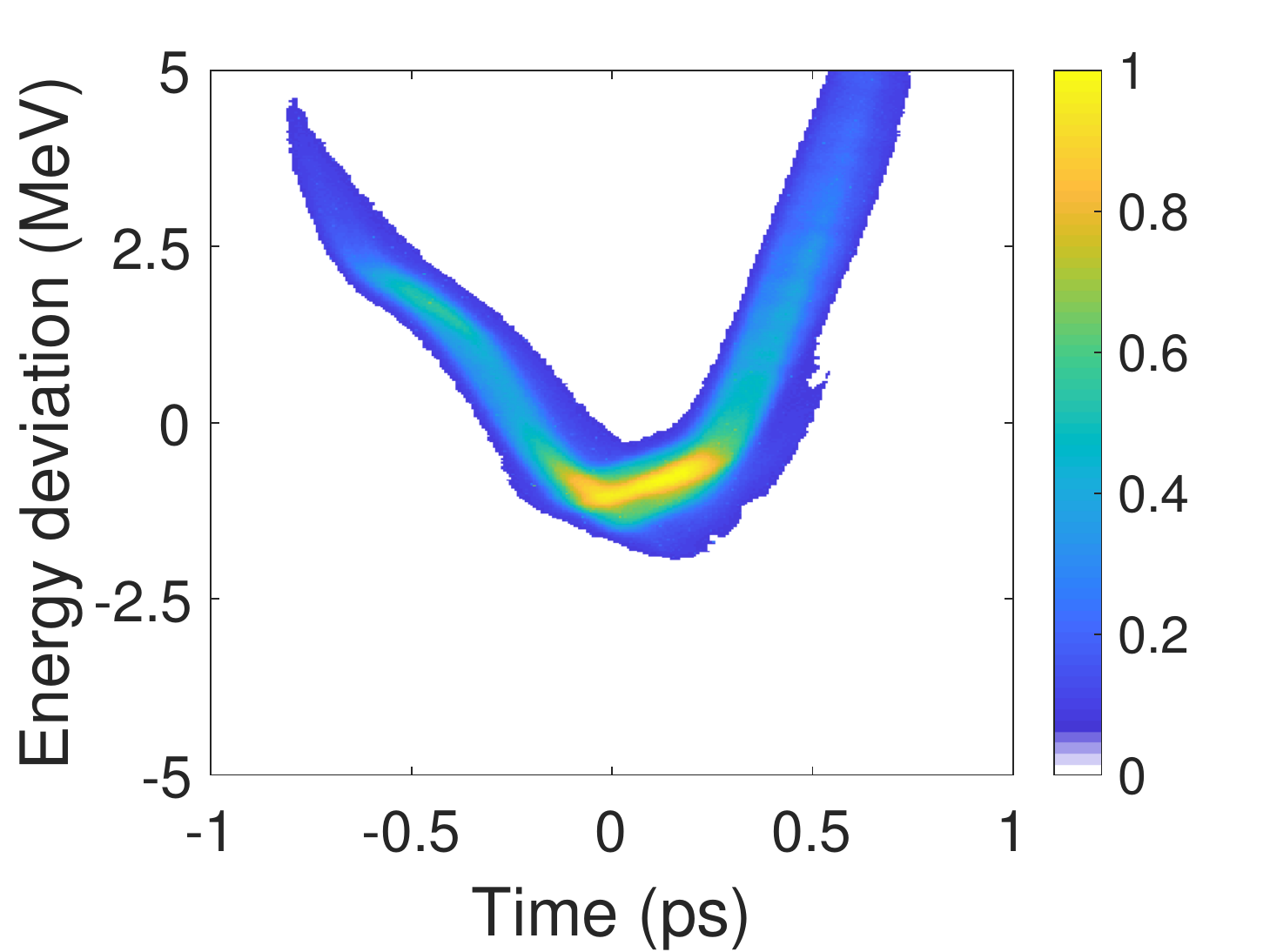}}
	\subfigure[]{\includegraphics*[width=0.47\linewidth]{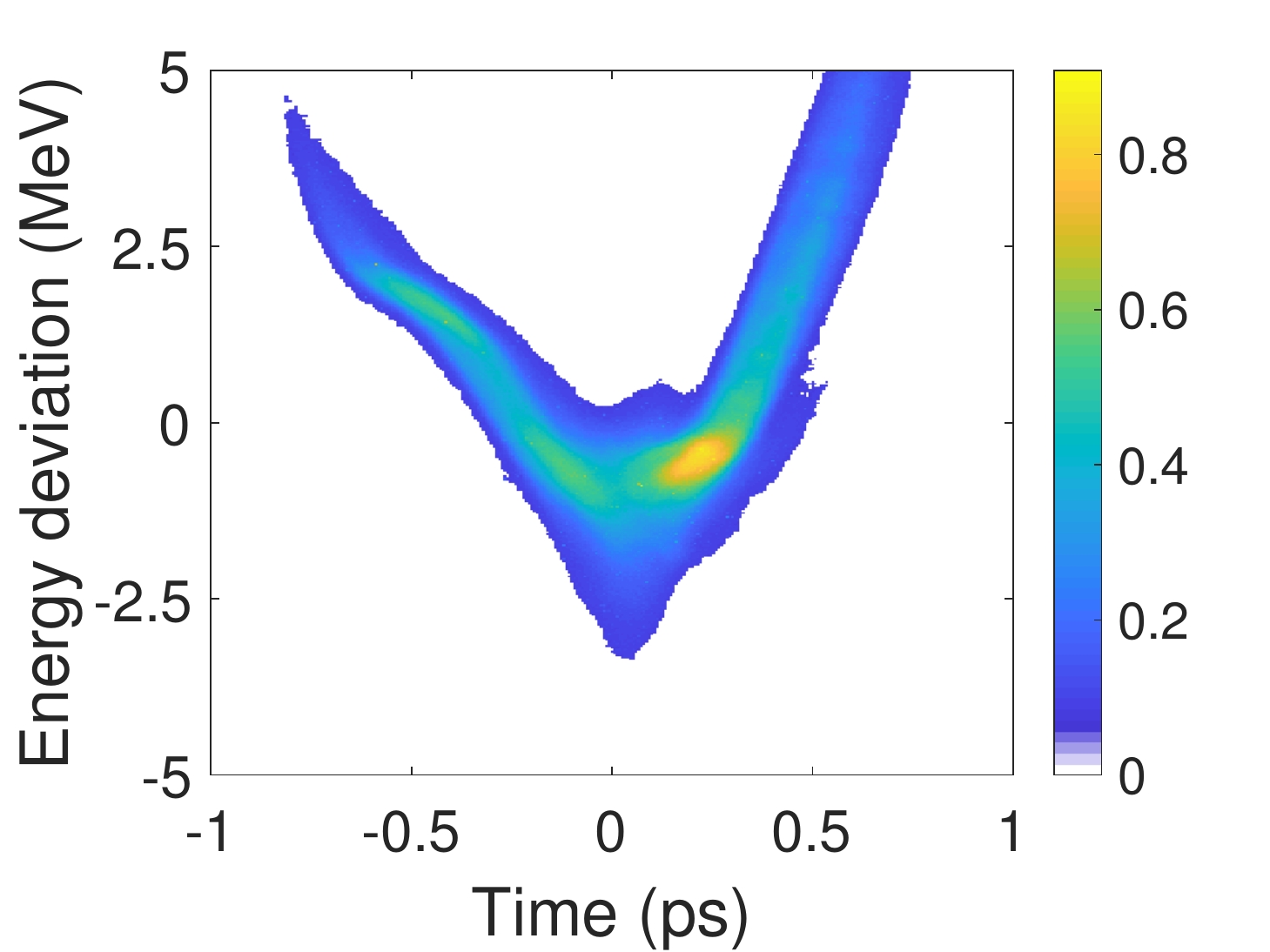}}
	\subfigure[]{\includegraphics*[width=0.47\linewidth]{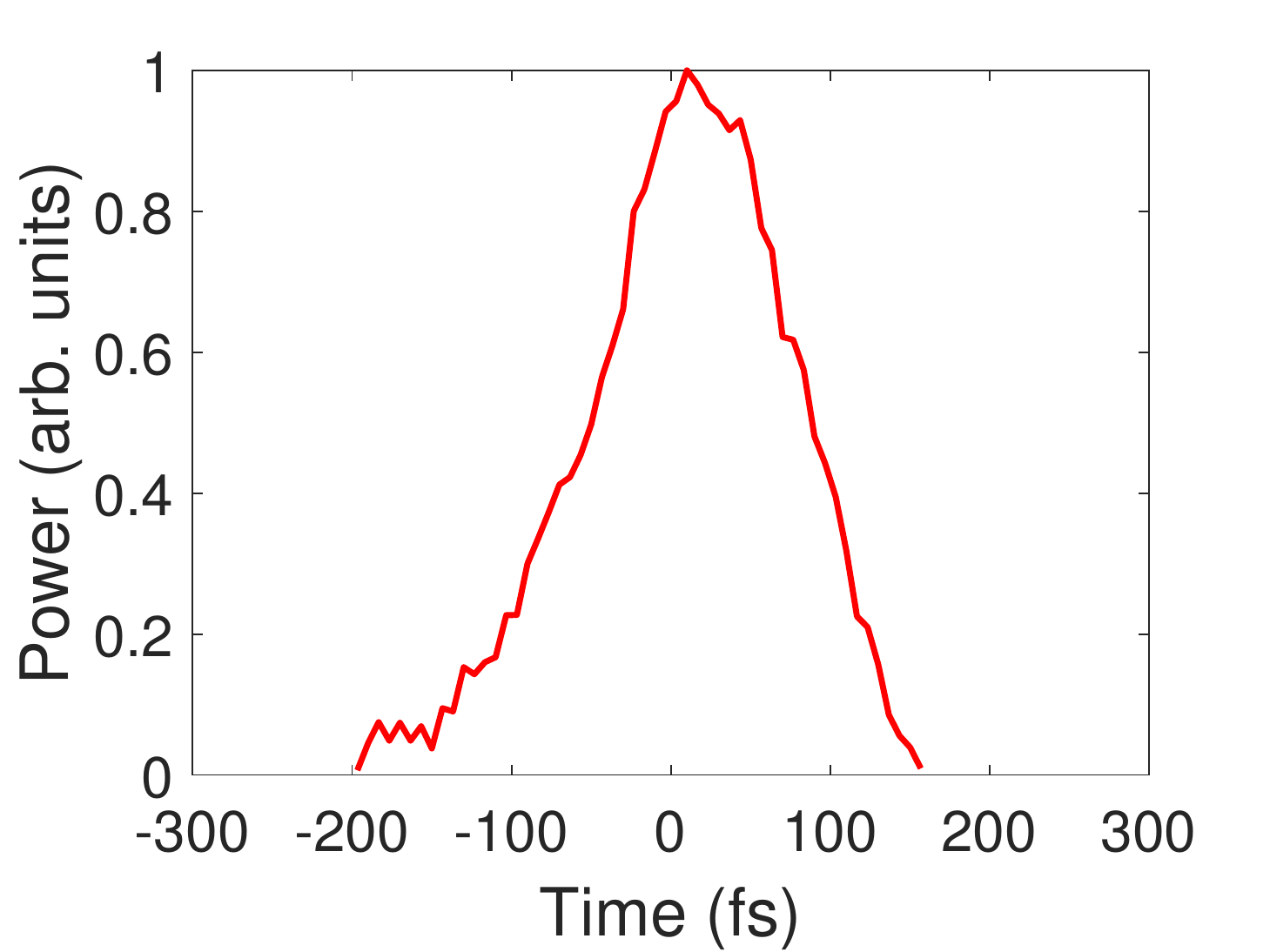}}
	\subfigure[]{\includegraphics*[width=0.47\linewidth]{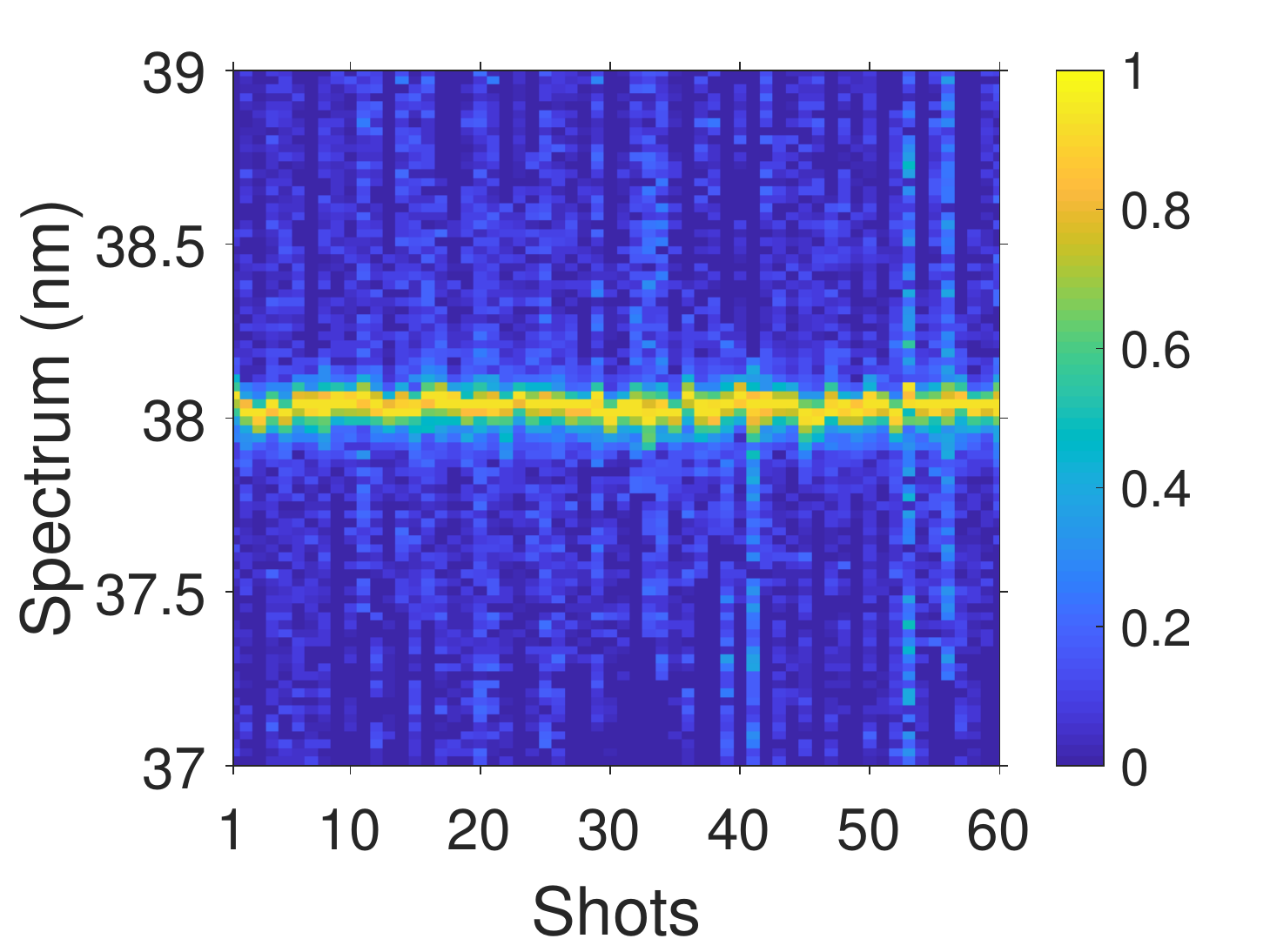}}
	\caption{Performance of the FEL lasing at 38 nm. The measured single-shot longitudinal phase space of the electron beam with the seed laser (a) turned off and (b) turned on. (c) Reconstructed FEL pulse temporal profile from the energy loss. Beam head is on the left in these plots.  (d) 60 consecutive single-shot spectra.}
	\label{tp}
\end{figure}

To verify the capability of the self-modulation scheme for FEL lasing, the other undulator segments of the radiator are used to further amplify the coherent radiation. The resonance condition of the radiator is maintained at the 7th harmonic of the seed laser. This is the first time that the first stage of the SXFEL operates at the 7th harmonic. The measured gain curve along the radiator and one typical transverse distribution of the laser pulse are shown in Fig.\ \ref{Gain} (b). At the exit of the radiator, FEL pulses with a mean energy of 17 $\rm \mu J$ and an rms energy jitter of 5 $\rm \mu J$ are obtained. The maximum pulse energy can reach 27 $\rm \mu J$.  Fig.\ \ref{tp} (a,b) presents two typical longitudinal phase spaces of the electron beams measured by the X-band transverse deflecting structure (XTDS) section \cite{SMaterial} at the exit of the undulator section with seed laser turned off and on, respectively. The measured bunch length and peak current of the electron beam are 0.92 ps (FWHM) and 623 A, respectively. According to the FEL pulse reconstruction method proposed in \cite{behrens2014few}, the temporal profile of a typical FEL pulse was obtained based on the beam energy loss. The reconstructed FEL pulse with a pulse length of 153 fs (FWHM) is displayed in Fig.\ \ref{tp} (c). Fig.\ \ref{tp} (d) presents 60 consecutive single-shot spectra of the FEL pulses. Theoretically, for a 153-fs transform-limited pulse, the corresponding relative bandwidth is $\rm 0.37\times10^{-3}$. The average relative bandwidth (FWHM) of the 60 shots is $\rm 2\times10^{-3}$, which is mainly limited by the resolution of 0.05 nm of the spectrometer.

\begin{figure}[htp] 
	\centering 
	\subfigure[]{\includegraphics*[width=0.45\linewidth]{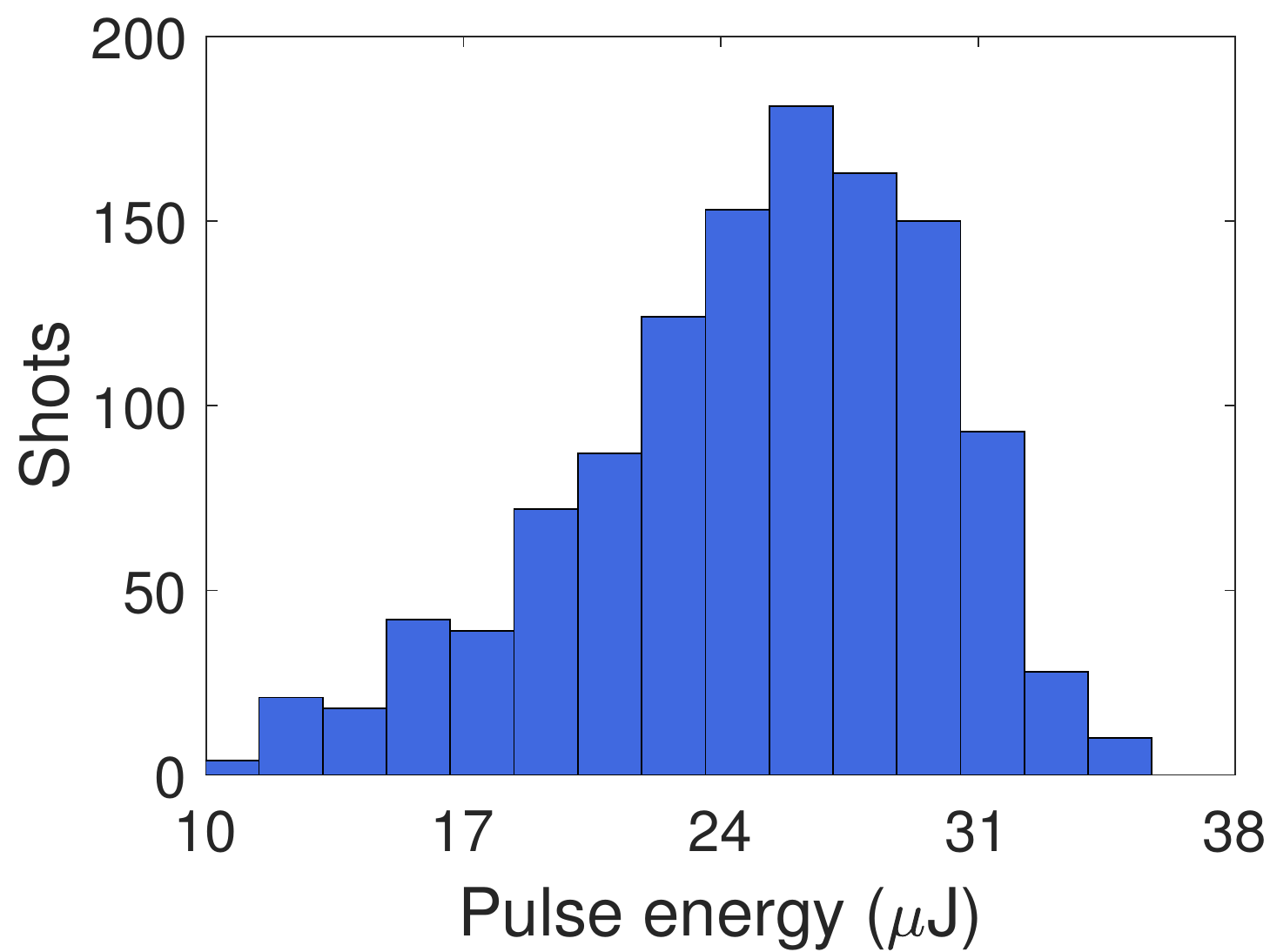}}
	\subfigure[]{\includegraphics*[width=0.45\linewidth]{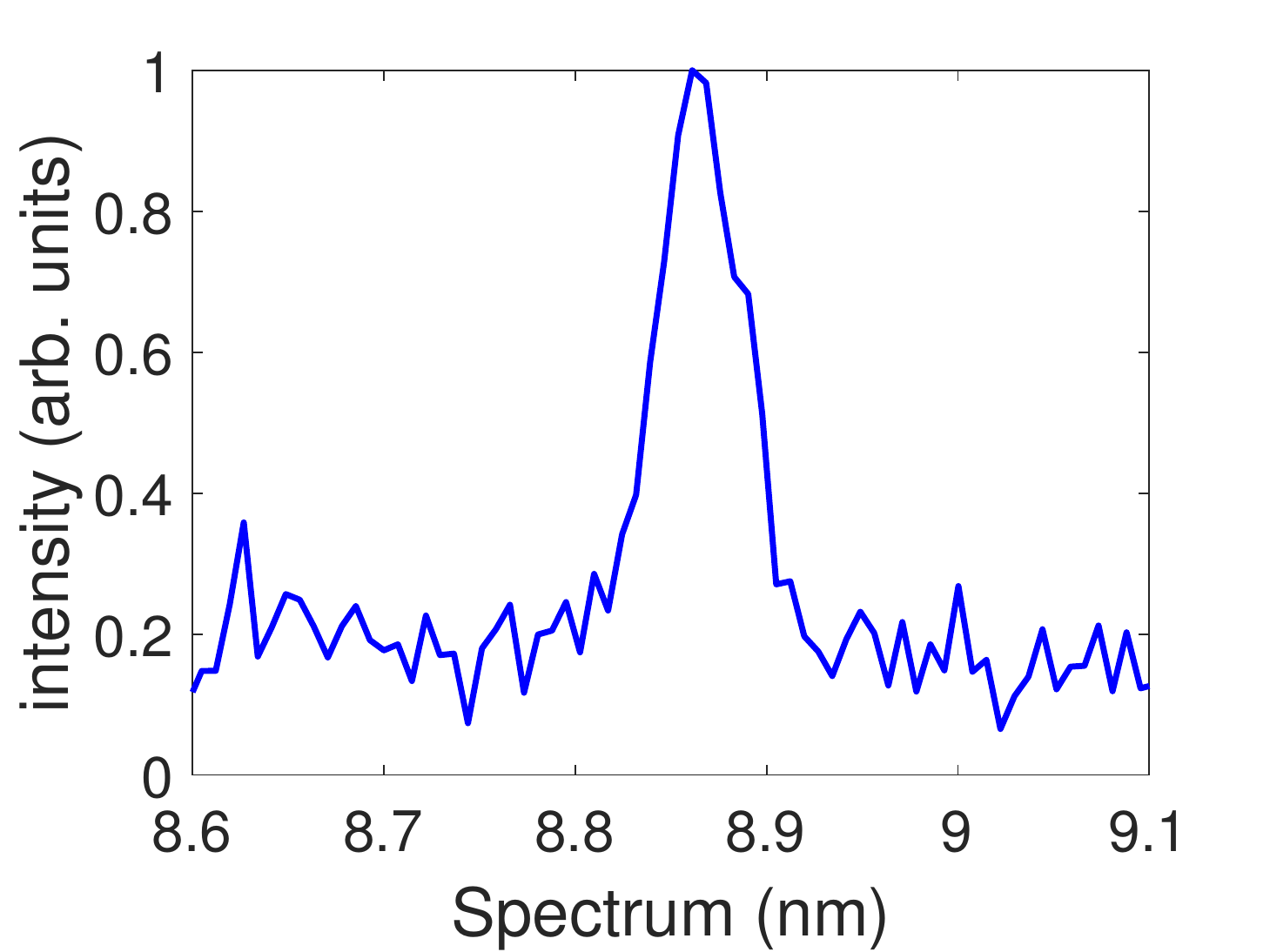}}
	\caption{Performance of the cascaded HGHG. (a) Pulse energy distribution of 1164 consecutive FEL pulses at 44.33 nm. (b) Spectrum of the cascaded HGHG at 8.87 nm (accumulated with integration over 20 shots).}
	\label{shots} 
\end{figure}

We further explored the feasibility of obtaining shorter wavelengths based on the cascaded HGHG scheme. The normal operation for SXFEL is a $\rm 6\times5$ cascading HGHG setup. Thus, in the experiment, we first changed the resonance of the radiator of the first stage from the 7th to the 6th harmonic by tuning the undulator gap and kept other parameters unchanged. Intense FEL radiation with a central wavelength of 44.33 nm was detected immediately. Fig.\ \ref{shots} (a) presents the measured pulse energy distribution for 1164 consecutive FEL pulses. The statistical results indicate that the average pulse energy is 25$\rm \mu J$, and the rms pulse energy jitter is 5 $\rm \mu J$.  After the first stage, the second stage of the SXFEL consists of a fresh bunch chicane, a modulator with a period of 55 mm, a dispersion section, and a radiator composed of six undulator segments with a period of 23.5 mm. Here, the second stage directly follows the parameters of the normal operation. The radiation generated from the first stage is shifted ahead to a fresh part of the electron beam by the fresh bunch chicane and serves as the seed laser in the following modulator. After interaction with the 44.33-nm FEL radiation and density modulation, the microbunched beam was sent to the radiator of the second stage whose resonance was tuned to the 5th harmonic of the first stage. Then, 8.87-nm coherent radiation with a pulse energy of approximately 0.5 $\rm \mu J$ was immediately detected, where the FEL gain mainly came from the first two undulator segments of the radiator, and its performance is very similar to that under the standard two-stage HGHG. The radiation spectrum accumulated with integration over 20 shots is presented in Fig.\ \ref{shots} (b), which is measured by a spectrometer with a resolution of 0.02 nm. The measured relative bandwidth (FWHM) is $\rm 5\times10^{-3}$. By carefully optimizing the beam orbit and focusing in the following undulator segments, larger FEL pulse energies can be expected. 

In summary, we have demonstrated a novel method to amplify laser-induced energy modulation through the self-modulation of a microbunched electron beam. Driven by the self-modulation, an electron beam with a laser-induced energy modulation as small as 1.8$\sigma_{E}$ was used to generate FEL radiation at the 7th harmonic in a single-stage HGHG scheme and the 30th harmonic in a cascaded HGHG scheme. In this case, a threefold increase in energy modulation amplitude is achieved. Theoretical analyses and further experiments \cite{SMaterial,yan2020observation} indicate that a more than fivefold enhancement in energy modulation amplitude, i.e.,  twenty-five-fold reduction in the peak power requirement of an external seed laser can be achieved through better control of the electron-beam envelope and orbit. This method paves the way for high-repetition-rate seeded FELs and storage-ring-based FELs.

Moreover, the self-modulation can be used to solve other critical problems. For example, the self-modulation can be employed by echo-enabled harmonic generation scheme \cite{feng2010pre, feng2015theoretical} or those seeded FEL schemes that require a high-power laser \cite{Hai_Xiao_2010,PhysRevAccelBeams.19.090701,PhysRevAccelBeams.20.010702} to achieve ultra-high harmonics. Besides, the low-peak-power seed laser allows for longer pulse duration and a larger transverse profile, so the self-modulation can greatly improve the stability of seeded FELs. In addition to conventional laser systems, high harmonic generation in gases is a promising extreme ultraviolet seeding source for FELs, but with low intensity \cite{lambert2008injection,labat2011high,ackermann2013generation}. Using the self-modulation to amplify the energy modulation induced by such sources would be a promising option for further frequency up-conversion \cite{doi:10.1080/09500340.2011.586475}. Therefore, the results presented here open up many possibilities for future seeded FELs.

This work was partially supported by the National Key Research and Development Program of China (Grant Numbers 2018YFE0103100, 2016YFA0401900) and the National Natural Science Foundation of China (Grant Numbers 11935020, 11775293).

\bibliography{mybibfile}

\end{document}